\documentclass[twocolumn]{aastex631}

\begin{document}

\title{Turbulent Energy Conversion Associated with Kinetic Microinstabilities \\in Earth's Magnetosheath}

\correspondingauthor{Harry C. Lewis}
\email{h.lewis21@imperial.ac.uk}
\affiliation{Department of Physics, Imperial College London, London, SW7 2BW, UK}

\author[0009-0002-2243-9092]{Harry C. Lewis}%
\affiliation{Department of Physics, Imperial College London, London, SW7 2BW, UK}

\author[0000-0002-5702-5802]{Julia E. Stawarz}%
\affiliation{Department of Mathematics, Physics and Electrical Engineering, Northumbria University, Newcastle upon Tyne, NE1 8ST, UK}%

\author[0000-0002-6276-7771]{Lorenzo Matteini}
\affiliation{Department of Physics, Imperial College London, London, SW7 2BW, UK}

\author[0000-0002-7419-0527]{Luca Franci}%
\affiliation{Department of Mathematics, Physics and Electrical Engineering, Northumbria University, Newcastle upon Tyne, NE1 8ST, UK}%

\author[0000-0001-6038-1923]{Kristopher G. Klein}
\affiliation{Lunar and Planetary Laboratory and Department of Planetary Sciences, University of Arizona, Tucson, Arizona 85721, USA}%

\author[0000-0002-0622-5302]{Robert T. Wicks}%
\affiliation{Department of Mathematics, Physics and Electrical Engineering, Northumbria University, Newcastle upon Tyne, NE1 8ST, UK}%

\author[0000-0002-6536-1531]{Chadi S. Salem}
\affiliation{Space Sciences Laboratory, University of California, Berkeley, California 94720, USA}

\author[0000-0002-7572-4690]{Timothy S. Horbury}
\author[0009-0008-4095-9175]{Joseph H. Wang}
\affiliation{Department of Physics, Imperial College London, London, SW7 2BW, UK}



\begin{abstract}

Plasma in the terrestrial magnetosheath is characterised by very weak particle--particle collisions, so kinetic microinstabilities are thought to be responsible for regulating the thermodynamics of the plasma. By exciting electromagnetic waves, these instabilities redistribute free energy in velocity space, moulding the velocity distribution function (VDF) into a lower energy state. In the high-beta magnetosheath, relatively small perturbations to the VDF can easily excite instabilities compared to in the low-beta inner heliosphere. Since magnetic fields cannot do work on the particles, electric fields mediate energy exchange between the electromagnetic field and the bulk fluid properties of the plasma. We investigate signatures of non-ideal energy conversion associated with turbulent fluctuations in the context of electron and ion temperature anisotropy--beta instabilities, utilising over 24 hours of data spread over 163 distinct intervals of in situ magnetosheath observations from Magnetospheric Multiscale (MMS). We find that average energy conversion into fluid flow is enhanced along instability boundaries, suggesting that turbulence is playing a role in how free energy is redistributed in the plasma. The work enables a quantification of the energetics which are associated with the role of kinetic microinstabilities in regulating collisionless plasma thermodynamics. This work provides insight into the open question of how specific plasma processes couple into the turbulent dynamics and ultimately lead to energy dissipation and particle energisation in collisionless plasmas.

\end{abstract}



\section{\label{sec:Introduction}Introduction}

Plasma turbulence is the predominant process by which fluctuation energy is transferred from injection at large scales to dissipation at microphysical scales. This phenomenon is known as the turbulent cascade and operates in a wide range of astrophysical plasma environments \citep{Matthaeus2011,Galtier2018,Stepanova2022}, such as (but not limited to) planetary magnetospheres and ionospheres \citep{Guio2021,Saur2021}, stellar coronae \citep{Cranmer2015, Zank2021}, stellar winds \citep{Carbone2012, Bruno2013, Verscharen2019, Adhikari2021}, the interstellar medium \citep{Fraternale2022, Linsky2022}, and beyond \citep{Subramanian2006, Evoli2011, Ruszkowski2023a}. Despite decades of intensive research, the nature of the kinetic-scale interactions which mediate dissipation in collisionless space plasmas are still poorly understood \cite[e.g.][]{Alexandrova2013, Kiyani2015, Chen2016, Sahraoui2020, Matthaeus2021, Smith2021, Schekochihin2022, Marino2023}. 

The magnetosheath is a natural laboratory for collisionless plasma turbulence \cite[e.g.][]{Alexandrova2008, Zimbardo2010, Matteini2016, Chen2017, Chhiber2018, Yordanova2020, Rakhmanova2021}, comprised of decelerated, heated, and compressed solar wind plasma which has been deflected around Earth's magnetosphere. Magnetosheath dynamics are influenced by solar wind turbulence as it is processed by the bow shock, as well as fluctuations driven by the shock itself \citep{Huang2017, Sahraoui2020, Rakhmanova2021}. In addition to strong fluctuations, the magnetosheath is host to complex velocity distribution functions (VDFs) which deviate significantly from a Maxwellian distribution \cite{Shuster2019}. Like the turbulent fluctuations, the shape of the VDF is influenced by the processing of the shock as well as fluctuations in the magnetosheath. Nonthermal VDFs are susceptible to instabilities, which scatter free energy in velocity space, driving the plasma to relax into a state closer to thermodynamic equilibrium and exciting electromagnetic wave modes \citep{Gary1993,Treumann1997}. 

In collisionless plasmas, internal energy needs to be redistributed via wave–particle interactions, which requires that free energy is used to grow the wave. There are various ways by which free energy can be scattered, but the common factor is that, since magnetic fields ($\mathbf{B}$) cannot do work on particles, energy transfer occurs via interactions with the electric field ($\mathbf{E}$). Energy exchange between the electromagnetic field and bulk flow energy is mediated by $\mathbf{j}\cdot\mathbf{E}$, where  $\mathbf{j}=\sum_s q_s n_s \mathbf{v}_s$ is the total current density, $q_s$ is the charge, $n_s$ is the number density, $\mathbf{v}_s$ is the bulk flow velocity, and $s$ denotes the species. \citep{Baumjohann1996}. Taking the contribution from the electron-frame electric field $\mathbf{E}'=\mathbf{E}+\mathbf{v}_e \times \mathbf{B}$ gives the Joule dissipation, $\mathbf{j}\cdot\mathbf{E}'$ \citep{Zenitani2011,Voros2019}, which is the collisionless analogue of Ohmic heating, $\eta j^2$. This term is important for collisionless dissipation  \citep{Birn2005,Birn2009,Zenitani2011a} because it encapsulates energy conversion related to non-ideal electric fields, which are capable of decoupling particle motion from $\mathbf{B}$. In contrast to $\eta j^2$, which is positive by definition, $\mathbf{j}\cdot\mathbf{E}'$ is reversible and therefore sign-indefinite, although space plasma observations typically find it to be positive on average \citep{Matthaeus2020}. Positive values correspond to transfer from fields to flow (loading), whereas negative values correspond to conversion from the flow to the fields (generating) \citep{Birn2005}. Within the magnetosheath, $\mathbf{E}'$ is dominated by $\mathbf{E}_{P_e}=-\frac{1}{n_e e}\nabla \cdot \mathbf{P}_e$ (the pressure term of generalized Ohm's law, where $\mathbf{P}_e$ is the electron pressure tensor and $e$ is the elementary charge) at scales accessible to spacecraft measurements \citep{Voros2019,Stawarz2021,Lewis2023}. A key question in the study of collisionless plasma turbulence and its associated dissipation is understanding the origin of both the positive and negative energy conversion signatures and how they may relate to specific fundamental plasma processes.

Prior studies have demonstrated that temperature anisotropy ($T_{s,\perp}/T_{s,\parallel} \ne 1$) is constrained by the parallel plasma beta $\beta_{s,\parallel}=2 n_s k_B T_{s,\parallel} \mu_0 /B^2$, where $B=||\mathbf{B}||$ is the magnetic field strength, and $T_{s,\parallel}$ and $T_{s,\perp}$ are the temperature parallel and perpendicular to the local $\mathbf{B}$, respectively. Prior studies of the solar wind and magnetosheath have observed this effect for the protons \citep{Gary1995, Gary1997, Hellinger2006, Maruca2012, Matteini2013, Maruca2018, Bandyopadhyay2022a} and electrons \citep{Gary2005, Xu2012, Zhang2018, Graham2021}. The constraint put on  $T_{s,\perp}/T_{s,\parallel}$ by $\beta_{s,\parallel}$ is thought to be linked to a number of different instabilities \citep{Gary1993}, for which functional forms have been developed to parameterise the growth rate contours as a function of dimensionless parameters $S_e$, $\alpha_e$ (electrons) and $a$, $\beta_{0}$ (ions). These curves are expressed as $T_{e,\perp}/T_{e,\parallel}=1+S_e\:\beta_{e,\parallel}^{-\alpha_e}$ \citep{Gary1996} and $T_{i,\perp}/T_{i,\parallel} = 1 + a(\beta_{i,\parallel}-\beta_0)^{-b}$ \citep{Hellinger2006}, for the electrons and ions, respectively. When plotted on the anisotropy–beta plane, these curves serve as boundaries of the accessible region of parameter space, indicating that rapidly-growing kinetic microinstabilities have a controlling influence on the shape of VDFs. Electron parameter space is bounded by the electron whistler and oblique electron firehose instabilities, whereas the ions are influenced by the ion cyclotron, ion mirror, parallel ion firehose and oblique ion firehose instabilities. The instability growth rate thresholds are derived using linear Vlasov theory under the assumption of perturbations to single bi-Maxwellian components for each species growing in a uniform background. This is in apparent contradiction with the nonlinear, nonuniform conditions characteristic of turbulence, however the local nonlinear timescales are expected to be much greater than linear growth rates \citep{Bandyopadhyay2022}, although this is not always true, especially in the solar wind \citep{Klein2019,Klein2021}. \citet{Bandyopadhyay2022} found that in extreme regions of anisotropy--beta parameter space this condition is reversed, providing a possible explanation as to why linear theory is applicable to turbulent conditions such as those in the magnetosheath.

The links between turbulence and small-scale instabilities have recently started to be explored for the protons, for example in hybrid simulations of Alfv\'enic turbulence \citep{Markovskii2019,Bott2021} and in the solar wind \citep{Opie2023}. However, this is still an open problem. The resonant interactions responsible for particle energisation have been the subject of detailed studies in recent years \cite[e.g.][]{Howes2017, Klein2017, Chen2019, Verniero2021, Afshari2021, Jiang2024}, however they are difficult to investigate statistically because of their purely kinetic nature and how complex it is to look at individual features in the VDF. Instead, we adopt a fluid approach using $\mathbf{j} \cdot \mathbf{E}'$ to assess the net transfer of energy to and from the particles, neglecting how the dissipation takes place in detail on a particle-by-particle basis. In this letter, we look for the first time at how this fluid heating step is organised with respect to the plasma microinstabilities, to highlight how turbulent fluctuations are interacting with the kinetic processes which regulate the plasma thermodynamics.

\section{\label{sec:Data}Data}

Magnetospheric Multiscale (MMS) \citep{Burch2016} is a multi-spacecraft mission launched in 2015 to study magnetic reconnection in the geospace environment \citep{Fuselier2016,Tooley2016}. The mission features four probes orbiting in a tetrahedral formation, containing identical instrument suites which measure electromagnetic fields \citep{Torbert2016a,Russell2016,Lindqvist2016,Ergun2016} and VDFs \citep{Pollock2016} with unprecedented measurement cadence.  In its highest-resolution ``burst''-mode, MMS samples the electromagnetic field at up to 8192 Hz and produces electron and ion VDFs at cadences of 0.03 s and 0.15 s respectively. These rapid particle measurements facilitate in-depth exploration of sub-ion scale turbulence that was not possible with prior missions.

Our dataset consists of 24.4 hours of burst-mode data from 163 intervals of continuous magnetosheath measurements. Intervals were selected which are free from large-scale inhomogeneities and are all the equivalent of multiple correlation lengths of the turbulent fluctuations in duration. The intervals range from around 3---43 minutes in length, with a mean of 9 minutes. The interval list is derived from that used by \citep{Stawarz2022}, with additions up until the end of the 2019--2020 MMS dayside season. The interval list is provided as a CSV file in the supplementary material. Intervals are aggregated to create a single time series of variables on the electron and ion measurement cadence, as appropriate. Moments of the VDFs are used to compute $\mathbf{j}=e n_e (\mathbf{v}_i - \mathbf{v}_e)$ where quasineutrality is used, $e$ is the elementary charge, and $\mathbf{v}_i$ is interpolated onto the electron timescale, since small-scale fluctuations in $\mathbf{j}$ are dominated by $\mathbf{v}_e$ \citep{Stawarz2021}. $\mathbf{E}'$ is calculated on the electron instrument cadence, which requires averaging $\mathbf{E}$ and $\mathbf{B}$ onto the same timescale as $\mathbf{j}$. Instantaneous local values of $\beta_{s,\parallel}$ and $T_{s,\perp}/T_{s,\parallel}$ are used, where the ion values are interpolated onto the electron cadence under the assumption that small-scale fluctuations in these parameters can be neglected. Research in the present study is conducted using measurements at the electron instrument frequency (33.33 Hz).

\begin{figure}
    \centering
    \includegraphics[width=\linewidth]{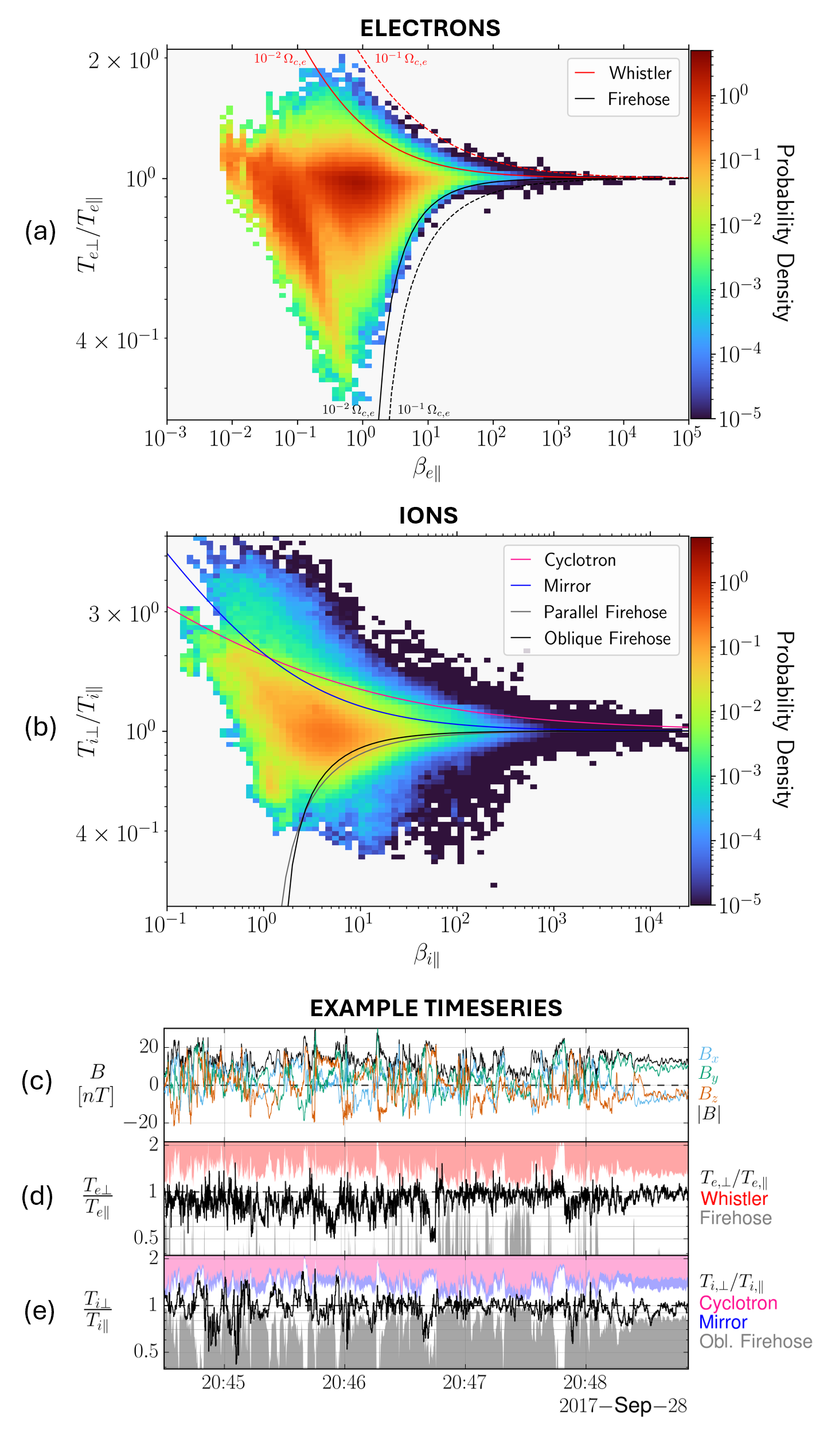}
    \caption{Probability density plotted on the anisotropy--beta plane for our MMS dataset, with an example time series. Panel (a) represents the electron plane, where solid and dashed lines correspond to $\gamma=10^{-2}\,\Omega_{c,e}$  and $\gamma=10^{-1}\,\Omega_{c,e}$ growth rate thresholds, respectively. Panel (b) shows the ion plane, with $\gamma=10^{-2}\:\Omega_{c,i}$ instability thresholds. Panels (c)--(e) illustrate motion through parameter space during 2017-09-28/20:44:30---20:48:51. Panel (c) shows the turbulent magnetic field. Panels (d) and (e) show the temperature anisotropy of the electrons and ions, respectively. Shaded areas represent instantaneous $\gamma=10^{-2}\,\Omega_{c}$ instability thresholds for each species.}
    \label{fig:Brazilplot}
\end{figure}

Figure \ref{fig:Brazilplot} panels (a) \& (b) show how microinstabilities constrain accessible regions of anistotropy--beta parameter space in our dataset. Electron and ion instability thresholds are plotted using the parametric curves defined in \cite{Gary2003} and \cite{Maruca2018}, respectively, wherein it is assumed that VDFs are anisotropic Maxwellians and that there is no anisotropy for one species when computing the growth curves of the other. Despite nonzero background non-Maxwellianity and correlation between $T_{e,\perp}/T_{e,\parallel}$ and $T_{i,\perp}/T_{i,\parallel}$ within our dataset, the plots in Figure \ref{fig:Brazilplot} demonstrate that the vast majority of probability density lies within instability growth rate thresholds of $\gamma = 10^{-2}\,\Omega_c$, as seen in previous studies \citep{Maruca2018, Graham2021}. Comparing Figures \ref{fig:Brazilplot} (a) and (b), the ions are typically able to deviate further from their respective instability growth rate curves, and crucially the ion probability density appears more diffuse across parameter space, irrespective of instability thresholds. In both panels, the greatest occupancy of parameter space is concentrated around $T_\perp/T_\parallel \sim 1$ and $\beta_{\parallel} \sim$ $10^{-1}$---$10^1$, indicating the typical conditions in the magnetosheath. Panels (c) to (e) show that turbulent fluctuations drive changes to both anisotropy and instability thresholds, indicating that a typical interval moves significantly around parameter space as opposed to filling an isolated region close to its average. To highlight this, an animation illustrating the trajectory of the example time series through anisotropy--beta space is provided in the supplementary material.

\section{\label{sec:Results}Results}

We investigate how energy conversion to/from the electromagnetic fields is organised with respect to kinetic microinstabilities by plotting $\langle | \mathbf{j}\cdot\mathbf{E}'|\rangle$ and $\langle \mathbf{j}\cdot\mathbf{E}'\rangle$ on the anisotropy--beta plane. To account for any large-scale variability between intervals, we normalise the time series of $\mathbf{j}\cdot\mathbf{E}'$ by its mean-removed root mean square (rms) $=\sqrt{\langle (\mathbf{j}(t)\cdot\mathbf{E}'(t) - \langle \mathbf{j}(t)\cdot\mathbf{E}'(t)\rangle)^2 \rangle}$ on an interval-by-interval basis. The data in these units represents an energy conversion fluctuation where a value of 1 corresponds to the characteristic fluctuation amplitude for the interval from which a given data point originates. Similar qualitative results are obtained based on unnormalised values of $\mathbf{j}\cdot\mathbf{E}'$. 

\begin{figure*}
\includegraphics[width=\textwidth]{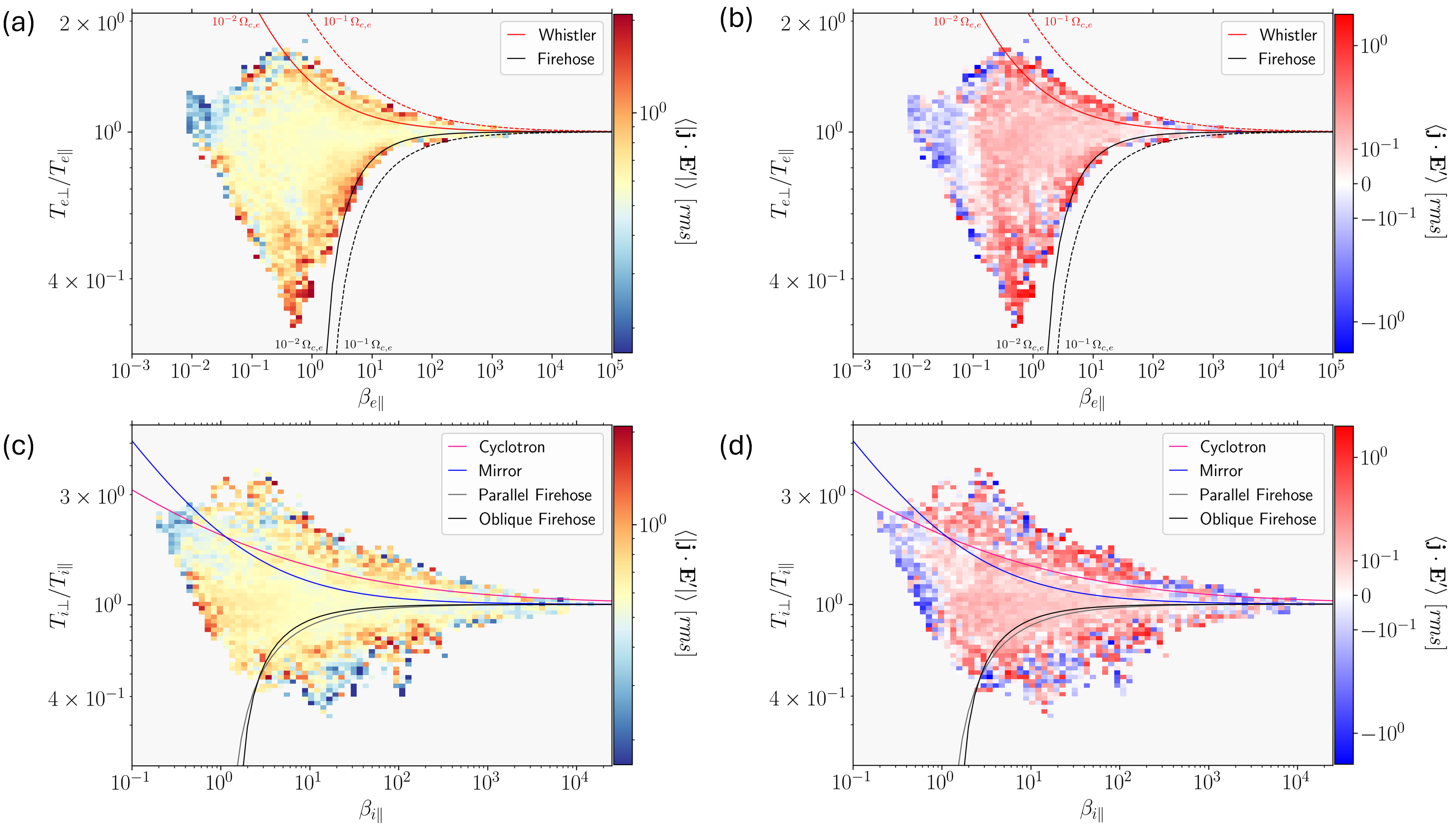}
\caption{\label{fig:jdotE}Average amplitude of energy conversion signals distributed on anisotropy--beta plane. (a) $\langle |\mathbf{j} \cdot \mathbf{E}' |\rangle$ in $T_{e,\perp}/T_{e,\parallel}$--$\beta_{e,\parallel}$ parameter space with instability thresholds of $\gamma=10^{-2} \, \Omega_{c,e}$ (solid line) and $\gamma=10^{-1} \, \Omega_{c,e}$ (dashed line). (b) $\langle \mathbf{j} \cdot \mathbf{E}' \rangle$ on the same plane. (c) $\langle |\mathbf{j} \cdot \mathbf{E}'| \rangle$ in $T_{i,\perp}/T_{i,\parallel}$--$\beta_{i,\parallel}$ parameter space with instability thresholds of $\gamma=10^{-2} \,\Omega_{c,i}$. (d) $\langle \mathbf{j} \cdot \mathbf{E}' \rangle$ on the same plane. In all panels, bins with fewer than 5 points in total or originating from fewer than 2 intervals are excluded.}
\end{figure*}

Figure \ref{fig:jdotE}(a) shows that there is enhanced $\langle | \mathbf{j}\cdot\mathbf{E}'|\rangle$ -- corresponding to a greater mean amplitude of energy conversion signals -- in regions susceptible to (oblique) electron firehose and electron whistler instabilities. Qualitatively, high $\langle | \mathbf{j}\cdot\mathbf{E}'|\rangle$ is most strongly concentrated around the firehose threshold, extending to $\gamma \le 10^{-2} \, \Omega_{c,e}$. In comparison, the enhancement associated with the whistler instability begins at $\gamma \ge 10^{-2} \, \Omega_{c,e}$. Interestingly, there is an enhancement along the low--$\beta_{e,\parallel}$ edge of the distribution for $T_{e,\parallel} > T_{e,\perp}$. Despite being partially obscured in the full set of intervals, closer investigation revealed that this feature is robustly present for different subsets of intervals. Given that this signature is not in the vicinity of the standard instability thresholds, the significance of enhanced $\langle | \mathbf{j}\cdot\mathbf{E}'|\rangle$ in this region of parameter space is less clear. It could be related to a possible instrumental bias shifting the PDF in anisotropy--beta space, but to determine this would require a separate systematic study. Figure \ref{fig:jdotE}(b) shows that mean signed energy conversion is positive across most of parameter space, and that it is particularly strong near the firehose and whistler instability thresholds. The centre of the distribution is weakly net-positive, accounting for the majority of overall energy conversion and suggesting that there is an average dissipation of electromagnetic energy into particle energisation. At low--$\beta_{e,\parallel}$, there are regions of negative average energy conversion which partially overlap with enhancements in $\langle | \mathbf{j}\cdot\mathbf{E}'|\rangle$.

There are interesting comparisons to draw with how energy conversion is organised across ion parameter space. Figure \ref{fig:jdotE}(c) shows that, as for electrons, $\langle | \mathbf{j}\cdot\mathbf{E}'|\rangle$ is enhanced along instability boundaries. However, the enhancement is most pronounced beyond the instability thresholds, suggesting that a faster growth rate is required to overcome the effects which are driving the ion VDFs into unstable regions. An enhancement on the low--$\beta_{i,\parallel}$ edge is also visible, albeit partially obscured as in the electron plot. This similarity may be expected because the time steps that are at $\beta < 1$ are mainly driven by large magnetic fields, appearing as low--$\beta_\parallel$ points on both species' planes. Figure \ref{fig:jdotE}(d) shows that $\langle \mathbf{j}\cdot\mathbf{E}'\rangle$ in ion parameter space is more sign-indefinite than the electrons beyond the instability thresholds, with average net conversion from fields to flow only being maintained in the stable centre of the distribution. 

\begin{figure*}
\includegraphics[width=.97\textwidth]{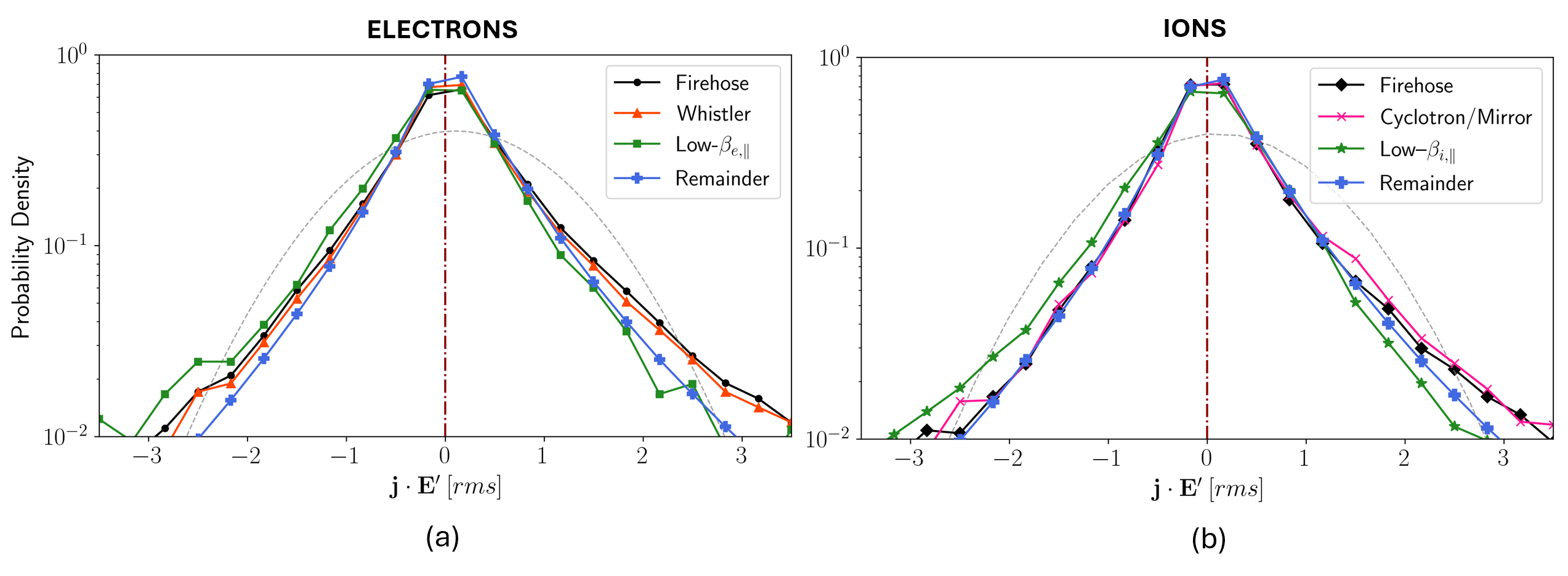}
\caption{\label{fig:region_pdfs} Probability density function of $\mathbf{j} \cdot \mathbf{E}'$ in each region for (a) the electrons and (b) the ions. Markers represent the centre of histogram bins. The dashed grey curve indicates a normal distribution with the same mean and standard deviation as the entire dataset.}
\end{figure*}

To investigate the statistical trends identified in anisotropy--beta parameter space, we manually select and group bins into regions of apparent enhanced $\langle | \mathbf{j} \cdot \mathbf{E}'|\rangle$ and investigate their properties. Certain regions have unclear boundaries (particularly those in ion parameter space), which introduces some variability, however the main features we highlight are not impacted by this. Note that the enhancements associated with the ion cyclotron and ion mirror instabilities have been treated as a single region. Figure \ref{fig:region_pdfs} shows the probability density function (PDF) in each manually-defined region for bins in the range $[-3.5, 3.5]$ rms units, within which the vast majority (98.10\%) of points lie. For context, 54.77\% of the dataset has $\mathbf{j}\cdot\mathbf{E}' > 0$ rms units, but only 10.10\% has  $\mathbf{j}\cdot\mathbf{E}' \ge 1$ rms units and 6.58\% has $\mathbf{j}\cdot\mathbf{E}' \le -1$ rms units. All regions, including the remainders, have nonzero skewness and high kurtosis. Panel (a) shows there is a slight enhancement in the negative tails of the electron PDF relative to the remainder, but a more significant enhancement on the positive side, leading to net positive $\mathbf{j} \cdot \mathbf{E}'$. The low--$\beta_{e,\parallel}$ region has diminished positive tail, but the heaviest negative tail of all the regions. Bins with $|\mathbf{j} \cdot \mathbf{E}'| \gtrsim 2$ rms units do not have enough counts to discern a reliable trend. Panel (b) demonstrates that ion instability regions are nearly identical to the remainder in the negative tail, but are enhanced on the positive side. The low--$\beta_{i,\parallel}$ region is reminiscent of its electron counterpart because they largely share similar points; the negative tails are heavier than positive, and the peak value is smaller. These trends will be assessed quantitatively in the next section.

\section{\label{sec:Discussion}Discussion}

\begin{table*}
    \begin{tabular}{|c|c|c|c|c|c|c|}
        \hline
        \centering
        Region & Data (\%)& $\langle \mathbf{j}\cdot\mathbf{E}'\rangle$& $\langle | \mathbf{j}\cdot\mathbf{E}'| \rangle$ & Variance & Skewness & Kurtosis\\\hline
        Electron firehose & 2.48 & 0.159 & 0.779 & 2.00 & 3.01 & 79.20\\
        Electron whistler & 0.56 & 0.150 & 0.721 & 1.76 & 5.25 & 164.99\\
        Electron low--$\beta_{e,\parallel}$ & 0.14 & -0.059 & 0.755 & 3.15 & 15.96 & 648.70\\\hline
        Electron remainder & 96.82& 0.104 & 0.583 & 0.97 & 1.89 & 94.50 \\\hline
        Ion firehose & 0.54 & 0.127 & 0.683 & 1.67 & 2.92 & 61.60\\
        Ion cyclotron/mirror & 0.44 & 0.155 & 0.713 & 2.23 & 5.55 & 224.16\\
        Ion low--$\beta_{i,\parallel}$ & 0.67 & -0.040 & 0.667 & 1.79 & 1.33 & 824.73\\\hline
        Ion remainder & 98.35& 0.106 & 0.587 & 0.99 & 2.09 & 85.51\\
        \hline
    \end{tabular}
    \caption{Mean net and amplitude of $\mathbf{j}\cdot\mathbf{E}'$, variance, skewness, and kurtosis by region.}
    \label{tab:region_stats}
\end{table*}

There are a number of processes involving energy conversion which may be related to kinetic microinstabilities. Turbulent fluctuations driving the plasma towards an instability threshold by distorting the distribution would be expected to give a positive signature. Growth of wave modes as a result of instabilities might be expected to give a negative signature. Redistribution of free energy in velocity space due to wave growth may not show up in $ \mathbf{j}\cdot\mathbf{E}'$. In our dataset, the presence of strong fluctuations in $ \mathbf{j}\cdot\mathbf{E}'$ along the unstable boundaries of the distribution is indicative of a link between turbulence and the underlying instability processes. Largely positive $\langle \mathbf{j}\cdot\mathbf{E}'\rangle$ across parameter space indicates that we predominantly observe the extraction of energy from turbulent fluctuations into the fluid flow. However, around the instability thresholds we see a mixture of net-positive and net-negative bins, particularly for the ions, which suggests we are also sampling part of wave growth due to instabilities. 

Table \ref{tab:region_stats} gives the average values of $\mathbf{j}\cdot\mathbf{E}'$ and $|\mathbf{j}\cdot\mathbf{E}'|$ in each manually-defined region as introduced in Section \ref{sec:Results}. The three electron regions represent 3.18\% of the total dataset, including 3.04\% associated with the electron firehose and whistler instabilities. $\langle\mathbf{j}\cdot\mathbf{E}'\rangle$ in the electron instability regions shows that the average energy conversion direction is from the fields to the fluid flow, and that the average value is increased compared to the remainder. This is consistent with electron VDFs being driven into unstable configurations by fluid heating associated with turbulent fluctuations. The low--$\beta_{e,\parallel}$ region has net energy conversion from the fluid flow into the fields, which suggests we are observing the scattering of free VDF energy into wave modes.  The fraction of data associated with ion regions is lower at 1.65\%, with only 0.98\% of measurements associated with enhancements in $\mathbf{j}\cdot\mathbf{E}'$ at the ion cyclotron and firehose instabilities. $\langle\mathbf{j}\cdot\mathbf{E}'\rangle$ in the ion regions is similar to electron regions, with net positive energy conversion along instability thresholds and net negative conversion at small $\beta_{i,\parallel}$.

Investigating the average amplitude of signals tells us about how much energy conversion is occurring regardless of direction, without positive and negative values cancelling each other out. The electron firehose region has the strongest $\langle|\mathbf{j}\cdot\mathbf{E}'|\rangle$ enhancement, with a 33.6\% increase compared to the remainder of the data. The electron whistler region has the smallest increase at 23.6\%, but this is still larger than the enhancement in any of the ion regions. The low--$\beta_{e,\parallel}$ region has a 29.5\% increase, showing that more energy conversion is occurring on average in this region than in the remainder, albeit on average from flow to fields. Compared to the remainder, the ion instability regions have enhanced $\langle|\mathbf{j}\cdot\mathbf{E}'|\rangle$, by 16.4\% (ion firehose) and 21\% (ion cyclotron), indicating that smaller energy conversion signatures are present around ion instability thresholds than at equivalent electron boundaries. This implies that energy conversion fluctuations to and from the ion VDFs are on average weaker where kinetic microinstabilities are growing or acting compared to the electron case, possibly as a result of the ions being more nonthermal due to their lower thermal velocity. Similarly to the electron equivalent, the low--$\beta_{i,\parallel}$ region has negative $\langle\mathbf{j}\cdot\mathbf{E}'\rangle$ and increased $\langle|\mathbf{j}\cdot\mathbf{E}'|\rangle$.

For electrons, MMS may not sample particle distributions fast enough to resolve all dynamics. This, and clear net positive $\langle\mathbf{j}\cdot\mathbf{E}'\rangle$, suggests we are preferentially observing the driving of electron VDFs into unstable regions. In contrast, ion dynamics take place on longer timescales, potentially indicating why we can observe fluctuations more clearly, explaining why $\langle|\mathbf{j}\cdot\mathbf{E}'|\rangle$ is enhanced but $\langle\mathbf{j}\cdot\mathbf{E}'\rangle$ is more sign-indefinite in unstable regions. Wave propagation may bias our statistics against observing negative $\mathbf{j}\cdot\mathbf{E}'$ in the electrons, as \citet{Svenningsson2022} found most electron whistler waves are observed away from their source region. Other instabilities may be the same, although the non-propagating electron firehose mode is expected to dominate compared to the propagating mode \citep{Cozzani2023}. However, waves resulting directly from ion instabilities have been observed in the solar wind \citep{Gary2016,McManus2024}. Additionally, due to Taylor's hypothesis \citep{Taylor1938}, which is frequently applied to the magnetosheath (and verified to be true for a subset of our intervals in \citet{Stawarz2022}), MMS generally observes spatial slices due to plasma advection rather than temporal slices due to evolution. To account for these effects, a simulation study into $\mathbf{j}\cdot\mathbf{E}'$ and kinetic microinstabilities would be valuable to track how the instability cycle is linked to turbulence as plasma evolves through the magnetosheath.

In this study, we focus on $\mathbf{j}\cdot\mathbf{E}'$ as the representation of non-ideal energy conversion between the electromagnetic fields and particle energy. However, this mechanism is not the only pathway towards dissipation \citep{Matthaeus2020,Pezzi2021}. In the statistical formulation of energy transfer channels, the pressure–strain interaction, $- (\mathbf{P}_s \cdot \nabla) \cdot \mathbf{v}_s$, encapsulates energy transfer between fluid flow and random thermal motion \citep{Yang2017, Yang2017a, Chasapis2018}. Since the collisionless energy transfer quantities are sign-indefinite yet are expected to lead to net dissipation on average, it would be interesting to understand how $\mathbf{j}_s\cdot\mathbf{E}$ and $- (\mathbf{P}_s \cdot \nabla) \cdot \mathbf{v}_s$ balance locally. However, $- (\mathbf{P}_s \cdot \nabla) \cdot \mathbf{v}_s$ is a difficult quantity to compute using observational data because the electron-scale pressure gradients are frequently smaller than can be measured by particle instruments \citep{Dahani2024}. Other examples of processes which can modify the internal energy are the divergence of the vector heat flux density \citep{Du2020}, and dissipation associated with an increase in entropy due to collisions \citep{Liang2019,Liang2020,Argall2022,Barbhuiya2024}. Even in weakly collisional plasmas, non-Maxwellian features in the VDF are subject to intraspecies collisions as a result of the dissipation of strong velocity-space gradients \cite[e.g.][]{Schekochihin2009,Pezzi2021}. Such entropy studies may provide a novel way of assessing the scattering of particles in the distribution function associated with the instabilities. Further studies may utilise future spacecraft missions or simulations to expand on the results we have presented here and investigate the link between energy conversion measures in the context of kinetic microinstabilities.

\section{\label{sec:Conclusion}Conclusion}

In this letter, we present a statistical investigation into non-ideal energy conversion associated with kinetic microinstabilities, revealing enhanced levels of $\mathbf{j}\cdot\mathbf{E}'$ close to temperature anisotropy--beta instability thresholds. In these regions, compared to the remainder of the distribution, energy conversion has increased amplitude and larger positive net value, indicating that we more often observe driving of VDFs into unstable configurations than scattering of free energy associated with the growth of wave modes. We find that the greatest enhancement in $\langle \mathbf{j}\cdot\mathbf{E}' \rangle$ and $\langle | \mathbf{j}\cdot\mathbf{E}'| \rangle$ occurs around the electron firehose instability boundary. In contrast, the largest increase in ion parameter space occurs beyond the ion cyclotron instability threshold. We also highlight an additional region of interest, present in both parameter spaces, along the low--$\beta_{\parallel}$ edge, at $T_{e,\perp}/T_{e,\parallel} < 1$ for the electrons and  $T_{i,\perp}/T_{i,\parallel} \sim 1$ for the ions. For both species, the low--$\beta_{\parallel}$ edge has net negative energy conversion (with a smaller absolute value than the net of the remainder), highlighting these regions as areas where significant energy is being imparted into electromagnetic activity. The origin of this effect is not yet understood. Despite accounting for a small percentage of the overall observations, we have shown that non-ideal energy conversion is enhanced when VDFs are configured such that they are susceptible to anisotropy-driven kinetic microinstabilities. We suggest that, while these processes are not responsible for the large-scale energy budget of the magnetosheath, the instability growth---excitation---scattering cycle is playing an important active role in how turbulent fluctuations influence the plasma thermodynamics.

There are some unresolved questions which warrant exploration in a further study. The relationship between kinetic microinstabilities and energy conversion parallel and perpendicular to the local $\mathbf{B}$ is interesting, because of the consequences for particle acceleration along field lines. In addition, there are extra sources of free VDF energy arising due to deviations from bi-Maxwellians which are unaccounted for by the anisotropy--beta instability thresholds \citep{Klein2018,Klein2019a}. Some work has been done to uncover how this feature is organised with respect to the anisotropy-driven kinetic microinstabilities \citep{Graham2021,Walters2023}, and future investigations should aim to link trends in the non-Maxwellianity to energy conversion. Finally, other energy conversion measures -- such as the pressure--strain interaction, and the contribution to fluid heating from different elements of $\mathbf{j}$ and various components of $\mathbf{E}$ -- may be treated with a similar analysis as in the present paper in order to uncover their relationship to the regulation of plasma thermodynamics.

\begin{acknowledgements}
    H.C.L. and J.E.S. are supported by the Royal Society University Research Fellowship No. URF\textbackslash R1\textbackslash 201286. L.F. was supported by the Royal Society University Research Fellowship No. URF\textbackslash R1\textbackslash 231710 and Science and Technology Facilities council (STFC) grant ST/W001071/1. H.C.L. and L.F. were supported by the International Exchange Grant No. IES\textbackslash R1\textbackslash 231251. K.G.K. was supported by NASA Grant 80NSSC19K0912. J.H.W. was supported by STFC studentship ST/X508433/1. This research was supported by the International Space Science Institute (ISSI) in Bern, through ISSI International Team project \#556 (Cross-scale energy transfer in space plasmas). J.E.S. acknowledges useful discussions as part of the ISSI International Team project \#23-588 (Unveiling Energy Conversion and Dissipation in Non-Equilibrium Space Plasmas). The authors wish to thank the entire MMS Team for their work on the mission. The data used in this study are publicly available through the MMS Science Data Center at \url{https://lasp.colorado.edu/mms/sdc/public/}. Elements of data analysis were performed using the pyrfu analysis package available at \url{https://github.com/louis-richard/irfu-python}. H.C.L. thanks colleagues at Imperial College London for valuable discussions.
\end{acknowledgements}


\bibliographystyle{aasjournal}



\end{document}